\documentclass[%
 reprint,
 amsmath,amssymb,
 aps,
prd,
nofootinbib
]{revtex4-1}

\newcommand\ee{\end{equation}}
\newcommand\be{\begin{equation}}
\newcommand\eea{\end{eqnarray}}
\newcommand\bea{\begin{eqnarray}}
\newcommand{\bV}{\mathbf{V}}

\newcommand{\bn}{\mathbf{n}}

\newcommand{\B}{\textrm{B}}
\newcommand{\F}{\textrm{F}}

\newcommand{\HH}{\mathcal{H}}

\newcommand{\tb}{\tilde{b}}
\newcommand{\tf}{\tilde{f}}

\newcommand{\rbr}[1]{\left(#1\right)}

\usepackage[normalem]{ulem} 
\usepackage{subfigure} 
\usepackage{graphicx}
\usepackage{dcolumn}
\usepackage{bm}
\usepackage{amsmath,amsfonts,amssymb,slashed,upgreek,color}
\usepackage{multirow}
\usepackage{cancel}
\usepackage{ulem}
\usepackage[citecolor=blue]{hyperref}
\usepackage{comment}
\usepackage{multirow}

\newcolumntype{C}[1]{>{\centering\arraybackslash}p{#1}}

\begin{document}


\title{Rescuing constraints on modified gravity using gravitational redshift \\ in large-scale structure}

\author{Sveva Castello}
\email{sveva.castello@unige.ch}
\author{Nastassia Grimm}
\email{nastassia.grimm@unige.ch}
\author{Camille Bonvin}
\email{camille.bonvin@unige.ch}
\affiliation{D\'epartement de Physique Th\'eorique and Center for Astroparticle Physics,
Universit\'e de Gen\`eve, Quai E. Ansermet 24, CH-1211 Gen\`eve 4, Switzerland}

\date{\today}

\begin{abstract} 
The distribution of galaxies provides an ideal laboratory to test for deviations from General Relativity. In particular, redshift-space distortions are commonly used to constrain modifications to the Poisson equation, which governs the strength of dark matter clustering. Here, we show that these constraints rely on the validity of the weak equivalence principle, which has never been tested for the dark matter component. Relaxing this restrictive assumption leads to modifications in the growth of structure that are fully degenerate with modifications induced by the Poisson equation. This in turns strongly degrades the constraining power of redshift-space distortions. Such degeneracies can however be broken and tight constraints on modified gravity can be recovered by measuring gravitational redshift from the galaxy distribution, an effect that will be detectable by the coming generation of large-scale structure surveys.
\end{abstract}

\pacs{Valid PACS appear here}
\maketitle

\section{Introduction} 
\label{sec:intro}

One of the main goals of large-scale structure surveys is to determine whether the laws of gravity at cosmological scales are consistent with General Relativity (GR). This is motivated by the fact that modified gravity theories are able to explain the observed accelerated expansion of the Universe at late time without a cosmological constant or a dark energy component (see e.g.~\cite{Koyama:2015vza, Joyce:2016vqv} for reviews), therefore providing a viable alternative to the standard $\Lambda$CDM cosmological model.  

Various theoretical frameworks have been developed in recent years to test deviations from GR. A first possibility is to adopt a model or a class of models (for example Horndeski models~\cite{Horndeski1974}) and to constrain the parameters of this model. A second possibility is to parameterize deviations from GR in a more phenomenological and model-independent way, at the level of Einstein's and conservation equations, and to constrain these deviations directly. In this paper, we concentrate on the second approach, which has been extensively used in galaxy clustering and weak lensing analyses~\cite{DES:2018ufa, Planck:2018vyg,eBOSS:2020yzd,Song:2010fg}.

Our goal is twofold: first, we will show that this phenomenological approach leads to tight constraints on deviations from GR with current data {\it only} when assuming that the theory of gravity preserves the weak equivalence principle (WEP). The WEP has been validated up to great precision for the particles of the Standard Model (see e.g.~\cite{Wagner:2012ui}), but it has never been tested for the unknown dark matter component. We will show that when allowing dark matter to violate the WEP, current data cannot distinguish between deviations in Einstein's equations and violations of the WEP due to unbreakable degeneracies. Secondly, we will demonstrate that, with future galaxy surveys such as DESI~\cite{DESI:2016fyo} and the SKA~\cite{Bull:2015lja}, we can rescue the constraints on gravity modifications and simultaneously test the validity of the WEP for dark matter. This is achieved by introducing a new observable into the game: a measurement of the distortion of time (also called gravitational redshift) from galaxy clustering.

\section{Modified gravity parameterization} 
\label{sec:MG}

First, we review the standard framework that is used to constrain GR in large-scale structure analyses. At late time, our Universe can be described by four fields: two metric perturbations, $\Phi$ and $\Psi$, which encode deviations from a homogeneous and isotropic geometry\footnote{We use the perturbed Friedmann metric: $\mathrm ds^2 = a^2[-(1+2 \Psi)\mathrm d \tau^2 + (1- 2 \Phi)\mathrm d \mathbf{x}^2]$, where $\tau$ denotes conformal time.}; and two fields that describe fluctuations in the matter content of the Universe, namely the matter density fluctuations, $\delta\rho$, and the matter peculiar velocity, $\mathbf{V}$. GR provides equations that relate these four fields.

A simple and generic framework to parameterize deviations from GR consists in modifying these equations with two phenomenological functions. More specifically, the functions $\mu$ and $\eta$ are introduced in the Poisson equation and the relation between the two gravitational potentials (see e.g.~\cite{Pogosian:2010tj})
\begin{align}
k^2\Psi &=-4 \pi G a^2 \mu(z,k)  \delta\rho\,,  \label{eq:poisson}\\
\Phi&= \eta(z,k) \Psi\,. \label{eq:anis}
\end{align}
The other two equations, namely the continuity equation and the Euler equation, are typically left unmodified in this approach. This implies that all constituents (standard matter, dark matter and photons)
behave in the same way under gravity, such that the WEP is preserved. One can consequently work in the so-called Jordan frame, where geodesics are not modified and all deviations from GR are encoded in the modified Einstein's equations~\eqref{eq:poisson} and~\eqref{eq:anis}. Within this theoretical framework, the functions $\mu$ and $\eta$ have been constrained by redshift-space distortions (RSD)~\cite{eBOSS:2020yzd} and by gravitational lensing~\cite{DES:2018ufa, Planck:2018vyg}, with a precision of 10-30 percent.~\footnote{The recent observation of the speed of gravitational waves~\cite{LIGOScientific:2017vwq} does not put direct constraints on $\mu$ and $\eta$. It is only in the case of specific models that these observations can be used to constrain the sign of these functions (see e.g.\ Table 1 in~\cite{Saltas:2018fwy}) or their amplitude for some subcases~\cite{Creminelli:2019kjy, Noller:2020afd}.}

However, there is no observational evidence motivating the assumption that dark matter obeys the WEP on cosmological scales and indeed a violation of the WEP arises in various modified gravity theories. This can occur at the fundamental level, within GR if dark matter obeys a dark (non-gravitational) fifth force~\cite{PhysRevLett.67.2926,1992ApJ...398..407G,Archidiacono:2022iuu,Barros:2018efl}, or in modified theories of gravity where dark matter and baryons are coupled differently to gravity~\cite{PhysRevLett.64.123,Gleyzes:2015pma,Gleyzes:2015rua}. Alternatively, it can appear as an effective violation, altering the way dark matter falls in a gravitational potential. This happens e.g.~in theories with screened modifications of GR~\cite{Hui:2009kc,Desmond:2020gzn} or with violations of the strong equivalence principle (Nordtvedt effect~\cite{PhysRev.169.1014}) if a large fraction of dark matter were made of compact objects~\cite{Clesse:2017bsw}, as well as in models with interacting dark matter and dark energy~\cite{Wands:2012vg,Asghari:2019qld}. 

When relaxing the validity of the WEP, allowing dark matter to behave differently than standard matter, then in the Jordan frame of baryons, the Euler equation for dark matter is modified and can generically be written as
\begin{align}
V_{\rm dm}'+V_{\rm dm}-\frac{k}{\mathcal H}\Psi= E^{\rm break}\,, \label{eq:euler_mod}
\end{align}
where $V_{\rm dm}$ denotes the dark matter velocity potential in Fourier space.
The exact functional form of $E^{\rm break}$ depends on the mechanism responsible for the violation of the WEP. In~\cite{Bonvin:2018ckp}, it was shown that, if dark matter is non-minimally coupled to a new degree of freedom such as a scalar or vector field, one generically obtains
\begin{align} 
E^{\rm break}=-\Theta (z, k) V_{\rm dm}+\frac{k}{\mathcal H}\Gamma (z, k)\Psi\,.\label{Eq:DefEbreak}   
\end{align}
The term proportional to $\Gamma$ encodes the strength of the fifth force propagated by the new degree of freedom on dark matter, whereas the term proportional to $\Theta$ is a friction term describing the impact of the new degree of freedom on the redshifting of the velocity. For specific models, these functions can be related to the fundamental parameters in the Lagrangian~\cite{Bonvin:2018ckp}.

Our goal is to determine how the phenomenological modifications can be constrained by large-scale structure surveys. Note that, in practice, in a large majority of models only one set of parameters appears: in modified gravity theories with universal coupling to all matter components, only $\mu$ and $\eta$ are relevant ($\Gamma$ and $\Theta$ are zero), whereas in models with a dark fifth force acting solely on dark matter, only $\Gamma$ and $\Theta$ are relevant ($\mu$ and $\eta$ are equal to one). However, since our goal is to determine if observations are able to distinguish between these two types of models, it is essential to include all parameters in the analysis and investigate whether observations can constrain them separately. Additionally, there are also classes of models encoding the most general scenario where all modifications are allowed, e.g.~\cite{Gleyzes:2015pma}.

Galaxy clustering can be used to probe the growth of density fluctuations. To determine how this growth is affected by deviations from GR and by a fifth force, we combine Eqs.~\eqref{eq:euler_mod},~\eqref{Eq:DefEbreak} and~\eqref{eq:poisson} with the continuity equation, to obtain an evolution equation for the dark matter density. For simplicity, we assume in the following that the growth and velocity of galaxies, $\delta_g$ and $V_g$, are driven by that of dark matter such that $\delta_g=b\,\delta=b\,\delta_{\rm dm}$ and $V_g=V=V_{\rm dm}$, where $b$ is the galaxy bias. In Appendix~A, we show that modifying these relations to include a fraction of baryons does not impact the results of our analysis. We obtain 
\begin{align}
\delta''+ \left(1+\frac{\HH'}{\HH}+\Theta\right) \delta'-\frac{3}{2}\frac{\Omega_{m,0}}{a}\left(\frac{\HH_0}{\HH}\right)^2 \mu\,(\Gamma+1)\,\delta=0\,, \label{eq:deltaevol_mod}
\end{align}
where a prime denotes derivatives with respect to $\ln a$. We see that the growth of density fluctuations, which can be probed with galaxy clustering, is directly sensitive to the functions $\mu, \Theta$ and $\Gamma$. On the contrary, galaxy clustering is not sensitive to $\eta$, which can however be constrained by weak lensing~\cite{DES:2018ufa}.
It is clear from Eq.~\eqref{eq:deltaevol_mod} that there is a complete degeneracy between $\Gamma$ and $\mu$. This reflects the fact that the clustering of dark matter can be enhanced in two ways: either by adding a fifth force acting on dark matter ($\Gamma>0$), or by increasing the depth of the gravitational potential associated to a given density distribution ($\mu>1$), which in turn increases the infall and clustering of dark matter. In addition, we expect from Eq.~\eqref{eq:deltaevol_mod} a further degeneracy between $\mu(\Gamma+1)$ and the parameter $\Theta$, which tends to slow down dark matter clustering through friction. 

To further simplify the analysis, it is common to assume that $\mu, \Theta$ and $\Gamma$ are independent of $k$, see e.g.~\cite{eBOSS:2020yzd}. This is motivated by the fact that in the quasi-static approximation the $k$-dependence can usually be neglected~\cite{Gleyzes:2015pma,Gleyzes:2015rua,Bonvin:2018ckp}. Moreover, one usually assumes that the modifications evolve proportionally to the background evolution of dark energy \cite{Planck:2015bue, Simpson:2012ra, Baker:2014zva}, i.e.~that they become relevant only during the phase of accelerated expansion of the Universe, such that
\begin{align}
&\mu(z) =1+\mu_0\, \Omega_\Lambda(z)/\Omega_{\Lambda,0}\,, \label{defmu0} \\
&\Theta(z) = \Theta_0\, \Omega_\Lambda(z)/\Omega_{\Lambda,0} \quad\mbox{and}\quad
\Gamma(z) = \Gamma_0\, \Omega_\Lambda(z)/\Omega_{\Lambda,0}\,.\nonumber
\end{align}

\section{Galaxy clustering observable} 
\label{sec:evol}

We now study how galaxy clustering can constrain the parameters $\mu, \Theta$ and $\Gamma$. Galaxy surveys map the distribution of galaxies and provide measurements of the galaxy number counts fluctuations
\be
\Delta\equiv\frac{N(\bn,z)-\bar N(z)}{\bar N(z)}\, ,
\ee
where $N$ is the number of galaxies per pixel detected in direction $\bn$ and at redshift $z$, and $\bar N$ denotes the average number per pixel. In the linear regime, the observable $\Delta$ is given by \cite{Bonvin:2011bg, Challinor:2011bk, Yoo:2009au}
\begin{align}
\Delta(\bn, z)&=b\,\delta-\frac{1}{\HH}\partial_r(\bV\cdot\bn)+\frac{1}{\mathcal H}\partial_r\Psi+\frac{1}{\mathcal H}{\dot \bV}\cdot \mathbf n \label{eq:Delta_rel}\\
&+\left(1-5s+\frac{5s-2}{\mathcal H r}-\frac{{\dot{\HH}}}{\mathcal H^2}+f^{\rm evol}\right)\mathbf V\cdot \mathbf n\nonumber\, ,
\end{align}
where $r$ is the comoving distance to the galaxies and a dot denotes derivatives with respect to conformal time. The parameter $\HH$ denotes the Hubble parameter in conformal time,  $s$ is the magnification bias and $f^{\rm evol}$ is the evolution bias.
The first term contains the effect of matter density perturbations, while the second term encodes the well-known redshift-space distortions~\cite{Kaiser:1987qv}~\footnote{Note that, technically, RSD are due to the velocity of baryons, as they are the component that emits light. However, it was shown in~\cite{Bonvin:2022tii} that, since baryons are confined in galaxies, RSD correlations are only sensitive to the velocity of the galaxy center of mass, which is dominated by the dark matter velocity. Here we equate these two velocities, $V_g=V_{\rm dm}$. In Appendix~A, we study the case where the center of mass velocity is determined by a superposition of baryons and dark matter, showing that the results are very similar.}. These two terms are significantly larger than the others and are the only ones that are measurable with current surveys. The third term on the first line encodes the contribution from gravitational redshift, which changes the apparent size of a redshift bin located inside a gravitational potential. This effect is directly proportional to the metric potential $\Psi$, since it is due to the distortion of time inside a gravitational well. The last two terms are Doppler effects. Note that all the terms beyond the first two have been called `relativistic distortions' in the literature, even though only gravitational redshift is truly a general relativistic effect.~\footnote{Gravitational lensing and other relativistic distortions contribute to $\Delta$ \cite{Bonvin:2011bg, Challinor:2011bk, Yoo:2009au}, but their impact on observations in the redshift range relevant for this work is negligible~\cite{Jelic-Cizmek:2020pkh,Euclid:2021rez}.}

The standard approach to extract information from $\Delta(\bn, z)$ consists in measuring the two-point correlation function $\xi \equiv \langle \Delta( \mathbf{n}, z) \Delta(\mathbf{n}', z') \rangle$. The first two terms in Eq.~\eqref{eq:Delta_rel} only  generate a monopole, quadrupole and hexadecapole in the correlation function. The contributions from relativistic distortions to these multipoles have been shown to be negligible~\cite{Jelic-Cizmek:2020pkh} and, in the flat-sky approximation, they can be written as
\begin{align}
\xi_0(z,d)&=\left[\tb^2(z) +\frac{2}{3}\tb(z)\tf(z)+\frac{1}{5}\tf^2(z)\right]\mu_0(z_*, d)\,,\nonumber\\   
\xi_2(z,d)&=-\left[\frac 43 \tf(z)\tb(z)+\frac 47 \tf^2(z)\right]\mu_2(z_*, d)\,, \nonumber\\
\xi_4(z,d)&=\frac{8}{35}\tf^2(z)\mu_4(z_*, d)\,,\label{eq:mult}
\end{align}
where $f(z)\equiv\frac{\mathrm d\ln(\delta)}{\mathrm d\ln(a)}$ is the growth rate of structure, $\tf(z)\equiv f(z)\sigma_8(z)$, $\tb(z)\equiv b(z)\sigma_8(z)$ and
\begin{align} \label{eq:mu2}
\mu_\ell(z_*,d)=\int\frac{\mathrm dk\,k^2}{2\pi^2} \frac{P_{\delta\delta}(k,z_*)}{\sigma_8^2(z_*)}j_\ell(kd) \, . 
\end{align}
Here, we have introduced a redshift $z_*$, chosen to be well before cosmic acceleration started such that the deviations in Eq.~\eqref{defmu0} vanish. The functions $\mu_\ell(z_*,d)$ are therefore fully determined by early-Universe physics, and are tightly constrained by CMB observations \cite{Planck:2018vyg}. The amplitude of the multipoles is then directly sensitive to the growth rate $\tf(z)= f(z)\sigma_8(z)$, which is affected by the parameters $\mu_0, \Theta_0$ and $\Gamma_0$ through Eq.~\eqref{eq:deltaevol_mod}. The multipoles can therefore be used to constrain these parameters.

The relativistic distortions, i.e.\ the last three terms in Eq.~\eqref{eq:Delta_rel}, have the particularity to generate odd multipoles in the correlation function~\cite{Bonvin:2013ogt, Croft:2013taa,Yoo:2012se} (or similarly in the power spectrum~\cite{McDonald:2009ud}). To detect these odd multipoles, it is necessary to cross-correlate two distinct populations of galaxies, for example a bright (B) and faint (F) population. The dominant odd multipole is the dipole, which is too small to be detected in current surveys~\cite{Gaztanaga:2015jrs}, but is expected to be robustly detected with DESI \cite{Bonvin:2015kuc,Beutler:2020evf} and SKA2~\cite{Bonvin:2018ckp,Saga:2021jrh}. This dipole depends differently on modified gravity parameters from the even multipoles, and thus plays a fundamental role in breaking parameter degeneracies. 

By combining Eqs.~\eqref{eq:euler_mod} and~\eqref{eq:Delta_rel}, we can write the relativistic distortions as 
\begin{align}
\Delta^{\rm rel}=\frac{E^{\mathrm{break}}}{\HH}+\left(\frac{5s-2}{\mathcal H r}-5s-\frac{{\dot{\HH}}}{\mathcal H^2}+f^{\rm evol}\right)\mathbf V\cdot \mathbf n\, .
\end{align}
We note that the gravitational redshift and Doppler terms combine into a contribution that is directly proportional to $E^{\mathrm{break}}$. The dipole is then given by
\begin{align}
&\xi_1(z,d)= \frac{\mathcal H}{\mathcal H_0} \nu_1(d,z_\ast)\left[5\tf\rbr{\tb_\B s_\F-\tb_\F s_\B}\rbr{1-\frac{1}{r\mathcal H}}\right. \nonumber\\
&+3\tf^2\Delta s\rbr{1-\frac{1}{r\mathcal H}}+\tf\Delta \tb\rbr{\frac{2}{r\mathcal H}+\frac{\dot{\mathcal H}}{\mathcal H^2}} \label{Eq:xi1}\\
&\left.+\Delta \tb\rbr{\Theta \tf-\frac 32 \frac{\Omega_{m,0}}{a}\frac{\HH_0^2}{\HH^2}\Gamma\,\mu\, \sigma_8(z)}\right]-\frac 25\Delta \tb\,\tf\,\frac{d}{r}\mu_2(d,z_\ast)\,, \nonumber
\end{align}
where $\Delta \tb=\tb_\B-\tb_\F$, $\Delta s=s_\B-s_\F$, and  
\begin{align}
\nu_1(z_\ast,d)&=\int\frac{\mathrm dk\,k}{2\pi^2} \mathcal H_0\frac{P_{\delta\delta}(k,z_\ast)}{\sigma_8^2(z_\ast)}j_1(kd)\,.
\end{align}
Here, we have assumed that $f^{\mathrm{evol}}=0$ for simplicity. This parameter will be directly measurable from the data~\cite{Wang:2020ibf}.
The last term in Eq.~\eqref{Eq:xi1} proportional to $\mu_2$ arises from the wide-angle contribution~\cite{Bonvin:2013ogt}. We note that $\mu, \Theta$ and $\Gamma$ enter the dipole in two ways: first, through their impact on $\tf$ (as in the even multipoles), but also directly through the first term in the last line of Eq.~\eqref{Eq:xi1}, arising from the breaking of the WEP.

\section{Current and future constraints on modified gravity} 
\label{sec:current}

\begin{figure}
\subfigure{\includegraphics[height=4.3cm]{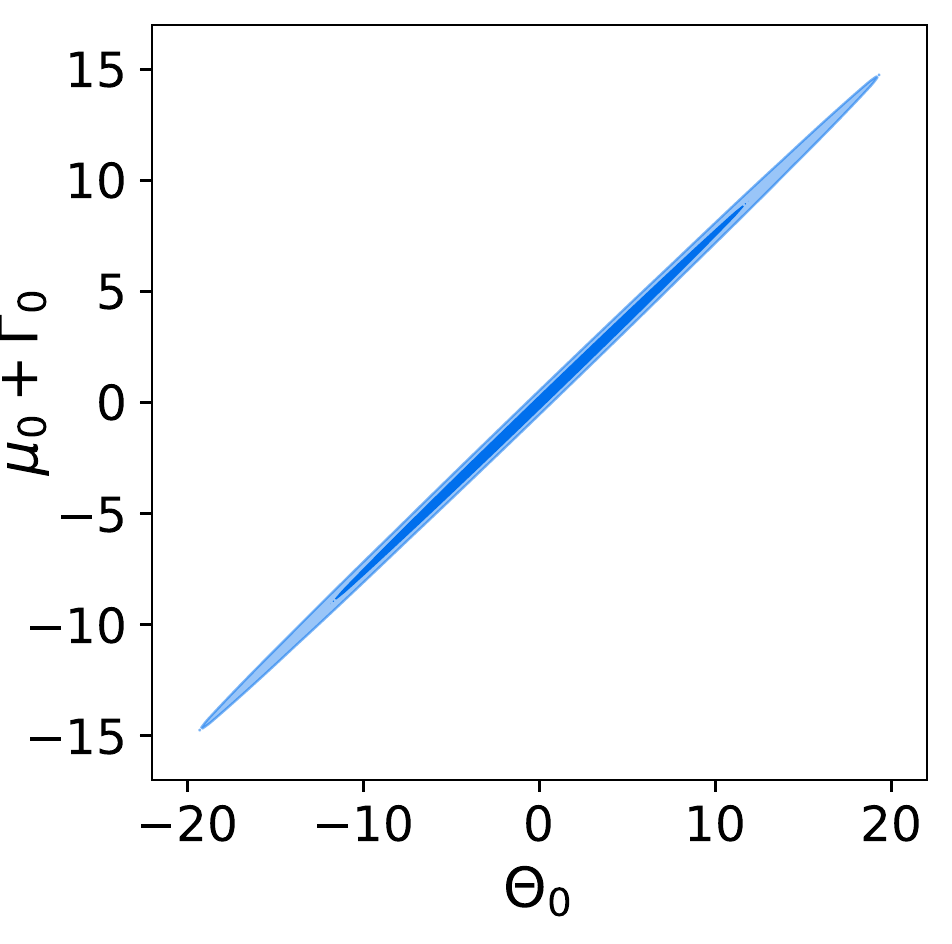}} \hspace{-0.2cm}
\subfigure{\includegraphics[height=4.3cm]{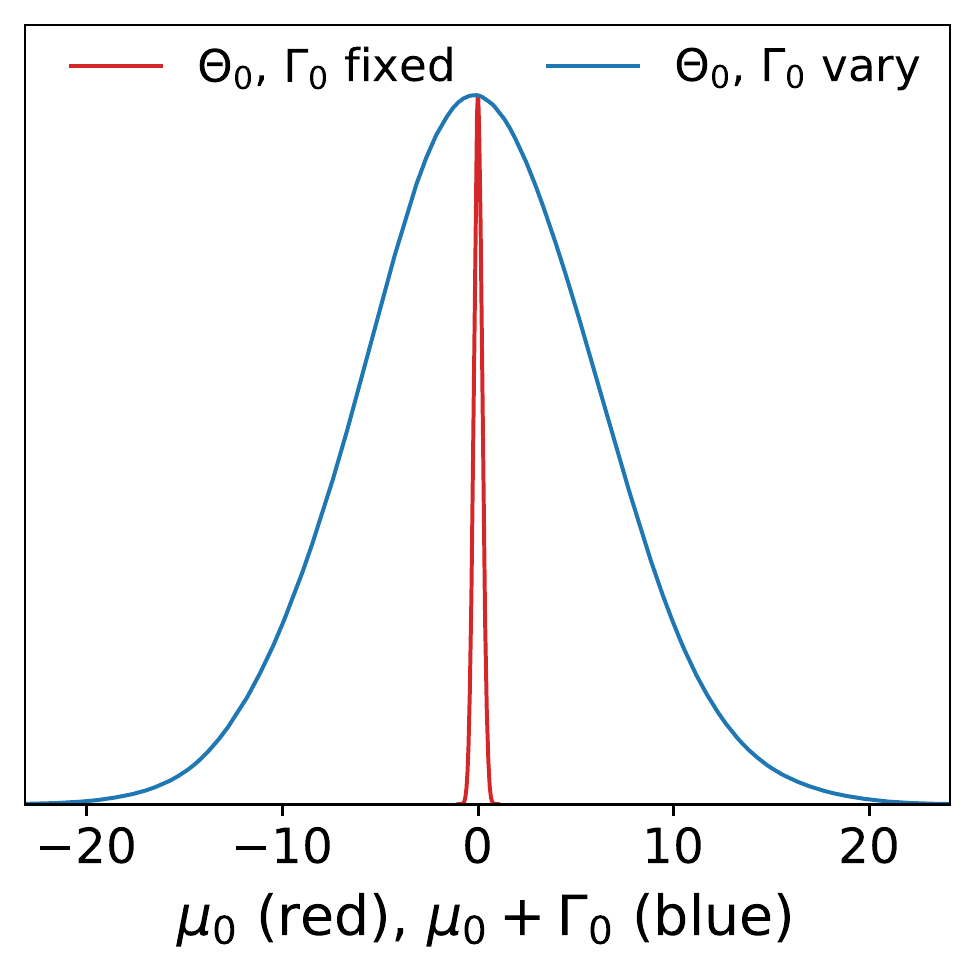}}
    \caption{In blue, we show constraints on $\Theta_0$ and $\Gamma_0+\mu_0$ from the RSD measurements of $\tilde f_i$ given in~\cite{eBOSS:2020yzd}. The red line corresponds to the much tighter constraint on $\mu_0$ under the restrictive assumption that the WEP is valid.}
   \label{Fig:CurrentConstraints}
\end{figure}

Given the expressions for the even and odd multipoles of the galaxy correlation function, we now investigate their constraining power on the parameters $\mu_0$, $\Theta_0$ and $\Gamma_0$. We first only include the even multipoles, as in standard large-scale structure analyses, and study how the constraints on $\mu_0$ are degraded if one does not impose the validity of the WEP. We then add the dipole and show that it allows us to recover tight constraints on all three parameters.

\begin{table}[t]
\centering
\caption{The RSD constraints on ${\mu_0}$, restricted to models where the WEP is valid, are degraded into much wider bounds on ${\mu_0 + \Gamma_0}$ when dropping this requirement.}
\begin{tabular}{cccc}
                                           & ~SDSS-IV~ & ~DESI~ & ~SKA2~ \\ 
                                           \hline \hline
$\sigma_{\mu_0}$ (restricted to WEP validity)              &   0.21 & {0.02}  & 0.004          \\ 
$\sigma_{\mu_0 + \Gamma_0}$ (no assumption on WEP) &  6.05 & {0.63} &  {0.062}       \\
\hline
\end{tabular} 
 \label{Tab:Constraints}
\end{table}

\subsection{Specifications for the analysis}

We consider both current data from SDSS-IV (including SDSS, BOSS and eBOSS), and future data expected from the coming generation of large-scale structure surveys. We focus on two future catalogues: the Bright Galaxy Sample (BGS) of DESI, which will observe 10 million galaxies up to $z=0.5$, and the SKA phase 2, which will observe close to a billion galaxies up to $z=2$. The survey specifications are taken from~\cite{DESI:2016fyo} and~\cite{Bull:2015lja} respectively. The fiducial cosmology is fixed to the latest \textit{Planck} values~\cite{Planck:2018vyg}. Note that we keep the background parameters fixed, as done in~\cite{eBOSS:2020yzd} for computing the constraints on~$\mu_0$, in order to obtain a fair comparison with current analyses. We let the bias vary according to the fitting functions given in~\cite{DESI:2016fyo, Bull:2015lja}, and marginalise over it: $b_{\rm BGS}=b_0\,\delta (0)/\delta(z)$ for DESI, involving one free parameter with fiducial value $b_0=1.34$; and $b_{\rm SKA}=b_1 \exp(b_2z)$ for SKA2, involving two free parameters with fiducial values $b_1 = 0.554$ and $b_2 = 0.783$. We choose $z_*=10$ and fix the minimum separation $d_{\rm min}=20$\,Mpc$/h$, such that non-linear effects are negligible~\cite{Bonvin:2020cxp}. We include shot noise and cosmic variance in the variance of the multipoles (see Appendix~C of~\cite{Bonvin:2018ckp}) and account for cross-correlations between different multipoles.  

\subsection{Constraints from even multipoles}

We first use the measurements of $\tilde{f}_i=f(z_i)\sigma_8(z_i)$ from the even multipoles of SDSS-IV (see Table~3 in~\cite{eBOSS:2020yzd}) to constrain $\mu_0, \Theta_0$ and $\Gamma_0$. When the WEP is enforced, we can directly translate the constraints on $\tilde{f}_i$ into constraints on $\mu_0$, and we find $\sigma_{\mu_0}=0.21$.\footnote{This is comparable to the value $\sigma_{\mu_0}=0.25$ obtained in~\cite{eBOSS:2020yzd} when combining RSD and weak lensing.} On the other hand, when dropping the assumption that the WEP is valid, the full degeneracy between $\mu_0$ and $\Gamma_0$ implies that only the sum $\mu_0+\Gamma_0$ (and not $\mu_0$ alone) can be constrained by RSD measurements. We therefore calculate a Fisher matrix for the parameter space $\{\mu_0 + \Gamma_0, \Theta_0\}$. The joint constraints are shown in Fig.~\ref{Fig:CurrentConstraints}. We see a strong degeneracy between $\mu_0 + \Gamma_0$ and $\Theta_0$, which is due to the fact that $\Theta_0$ slows down the growth of structure, while $\mu_0+\Gamma_0$ accelerates it. As a consequence, the marginalised constraint on $\mu_0+\Gamma_0$ is very large: $\sigma_{\mu_0+\Gamma_0}=6.05$, i.e.\ 30 times larger than the original constraints on $\mu_0$.

The same degeneracy affects constraints from future surveys. Our results are summarized in Table \ref{Tab:Constraints}, again showing a significant degradation of the constraints when allowing for a violation of the WEP. Hence, even multipoles are able to provide tight constraints on $\mu_0$ only under the restrictive assumption that dark matter obeys the WEP. In other words, if future surveys detect deviations from $\Lambda$CDM in the growth of structure, we will not be able to distinguish whether they are due to a modification of gravity or to a dark fifth force acting on dark matter.

Finally, let us mention that we have assumed $\mu_0$, $\Gamma_0$ and $\Theta_0$ to be scale-independent. Allowing for a scale-dependence would not break the degeneracies, unless when considering very specific models where $\mu_0$ has a known scaling (such as in $f(R)$ gravity models, see~\cite{Giannantonio:2009gi,Hu:2013aqa}) different from the one of $\Theta_0$ and $\Gamma_0$.

\begin{table}[t]
\centering
\caption{Forecasted constraints on $\mu_0+\Gamma_0$, the individual parameters $\{\mu_0$, $\Gamma_0$, $\Theta_0\}$ and the best-measured eigenvector $\lambda_1$ when adding the relativistic dipole to RSD measurements.}
\begin{tabular}{cccc}
                           & ~~DESI~~ & ~~SKA2~~  & ~~SKA2~with baryons~~ \\ \hline \hline
$\sigma_{\mu_0+\Gamma_0}$ & {0.60} & {0.062} &  0.077\\  \hline
$\sigma_{\mu_0}$    & {1.79}  & {0.147}  & 0.147      \\ 
$\sigma_{\Gamma_0}$ & {1.84}   & {0.158} & {0.186}      \\
$\sigma_{\Theta_0}$ & {0.73}  & {0.079} & {0.093} \\ \hline 
$\sigma_{\lambda_1}$ & {0.01} & {0.002} & {0.002} \\
\hline 
\end{tabular}  \label{Tab:Constraints2}
\end{table}

\begin{figure}
    \centering
    \includegraphics[width=0.46\textwidth]{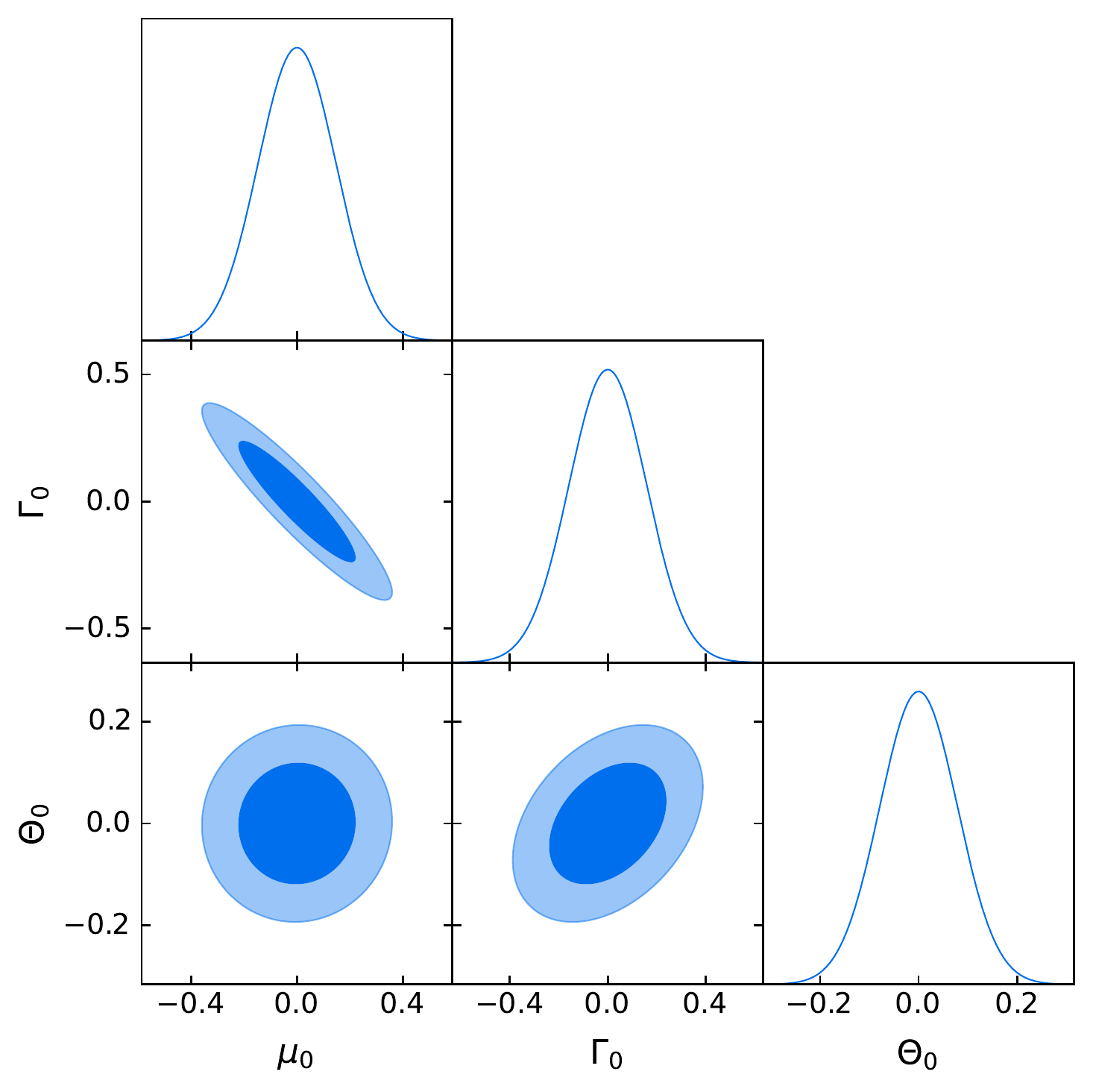}
    \caption{Forecasted constraints on $\mu_0$, $\Theta_0$ and $\Gamma_0$ for SKA2, combining the even multipoles and the dipole.}
    \label{fig:SKAcontourplot}
\end{figure}

\subsection{Adding the dipole as a \textit{Deus ex machina}}

We now combine the even multipoles and the dipole. We fix the bias difference to the value measured in BOSS for luminous red galaxies, $\Delta b=1$ \cite{Gaztanaga:2015jrs}, and let the free bias parameters vary separately for the bright and faint galaxy populations. The magnification bias is computed for the two populations using a Schechter function for the luminosity function~\cite{Jelic-Cizmek:2020pkh}, see Appendix B for detail.  

The marginalised constraints are presented in Table \ref{Tab:Constraints2} and the joint constraints for SKA2 are plotted in Fig.~\ref{fig:SKAcontourplot}. The bounds on $\sigma_{\mu_0 + \Gamma_0}$ are only marginally improved by adding the dipole. However, the dipole gives a decisive contribution by breaking the degeneracies, allowing us to constrain the three parameters individually. This arises from the gravitational redshift term $\partial_r \Psi$ that enters the dipole through $E^{\mathrm{break}}$ and leads to a further dependence on the parameters $\mu_0$, $\Theta_0$ and $\Gamma_0$ that is not present in the even multipoles. Gravitational redshift will therefore play a crucial role in future surveys, since it can distinguish between a modification of gravity, $\mu_0\neq0$, and a dark fifth force acting on dark matter, $\Gamma_0\neq 0$, $\Theta_0\neq0$. Moreover, in models where gravity is coupled differently to standard matter and dark matter, in which case all parameters have non-zero values (see e.g.~\cite{Gleyzes:2015pma}), gravitational redshift can constrain all modifications separately.

Since we are introducing two additional parameters, we expect that the bounds on $\mu_0$ are not as tight as in the RSD-only analyses with no violation of the WEP. Nevertheless, when diagonalizing the Fisher matrix, we always identify a combination of parameters that is very tightly constrained. For the baseline SKA2 analysis, this combination is $\lambda_1 = 0.62 \, \mu_0+0.62 \, \Gamma_0 -0.48 \, \Theta_0$, with an error $\sigma_{\lambda_1} = 0.002$, i.e.\ twice smaller than the original error on $\mu_0$ (see first line of Table~\ref{Tab:Constraints}). This indicates that the dipole contains additional information on deviations from $\Lambda$CDM with respect to the even multipoles. 

Finally, we have checked that the analysis holds if we add a fraction of baryons (obeying the WEP) to the density and velocity evolution (see Appendix A). The results are presented in Table~\ref{Tab:Constraints2}, and show that also in this case we obtain tight constraints on the individual parameters $\mu_0$, $\Gamma_0$ and $\Theta_0$.

\section{Conclusion} 
\label{sec:conclusion}
Our paper describes severe issues in testing gravity with upcoming large-scale structure surveys that previous work failed to see. Much effort has been devoted to designing and building these surveys, and it is therefore crucial to analyze the data in an optimal way to obtain robust constraints. Currently, constraints on modified gravity are obtained through RSD measurements, which directly probe the dynamical evolution of dark matter clustering, and are therefore sensitive to deviations in the Poisson equation, encoded in the parameter $\mu$.

We have shown that the clustering of dark matter is also affected by violations of the WEP. In particular, a fifth force acting on dark matter would enhance its clustering, whereas friction generated by an additional degree of freedom would suppress it. Since RSD are only sensitive to the growth rate of cosmic structures, they cannot distinguish between these different effects. Therefore, constraints on the parameter $\mu$ are completely spoiled by allowing for violations of the WEP for dark matter. Moreover, this loss of constraining power on $\mu$ propagates and spoils the constraints on $\eta$, i.e.\ on the anisotropic stress, since gravitational lensing can only measure the combination $\Sigma=\mu(1+\eta)$.

Luckily, as in all good plays, a \textit{Deus ex machina} rescues the situation: gravitational redshift in the galaxy number counts, which will be measurable with upcoming surveys. Since gravitational redshift is sensitive to the way photons escape from a gravitational potential, combining it with RSD (which probe the way dark matter falls in the potential) allows us to robustly constrain deviations from the WEP, and therefore to recover tight bounds on all parameters. 

\vspace{0.5cm}

\paragraph*{Acknowledgements.} We thank Levon Pogosian for interesting and useful discussions about the first version of this paper. CB thanks Francesca Lepori for useful discussions about the derivation of magnification bias. This project has received funding from the European Research Council (ERC) under the European Union’s Horizon 2020 research and innovation program (Grant agreement No.~863929; project title ``Testing the law of gravity with novel large-scale structure observables"). We acknowledge the use of the fftlog-python code written by Goran Jelic-Cizmek and available at \url{https://github.com/JCGoran/fftlog-python}. 


\section{Appendix A: Impact of a fraction of baryons}\label{app:baryons}

As shown in~\cite{Bonvin:2022tii}, the velocity that is relevant in RSD correlations is the velocity of the galaxy center of mass.
In our analysis, we have assumed that this velocity is completely determined by the velocity of dark matter, which is a good approximation since the galaxy mass is largely dominated by dark matter. Here, we study how the constraints change if we include the impact of baryons on the center of mass velocity. More precisely, we model the center of mass velocity as a weighted average of the contributions from the dark matter and baryonic components (see also~\cite{Umeh:2020cag}): $V=x V_{\rm dm}+(1-x)V_{\rm b}$, where $x\equiv \rho_{\rm dm}/\rho_{\rm m}$ is the fraction of dark matter inside a galaxy. Similarly, the matter density is given by $\delta=x \delta_{\rm dm}+(1-x)\delta_{\rm b}$.

This leads to a system of coupled differential equations
\begin{align} \label{eq:baryons1}
&\ddot{\delta}_{\rm dm}   +\HH(1+\Theta)\dot{\delta}_{\rm dm}\\
&-\frac{3}{2}\HH^2\Omega_m(z)\mu(1+\Gamma)\big[x\delta_{\rm dm}+(1-x)\delta_{\rm b} \big] =0\,,\nonumber\\
&\ddot{\delta}_{\rm b}   +\HH\dot{\delta}_{\rm b}-\frac{3}{2}\HH^2\Omega_m(z)\mu\big[x\delta_{\rm dm}+(1-x)\delta_{\rm b} \big] =0\,, \label{eq:baryons2}
\end{align}
which can be solved numerically using $\delta_{\rm dm}=\delta_{\rm b}$ as initial condition at $z_*$, when violations of the WEP are negligible, see Eq.~\eqref{defmu0}.

When including the impact of a fraction of baryons into our analysis, we expect a small degradation of the constraints. This is because only a part of the total matter (the dark matter component) is affected by the breaking of the WEP, leading to a factor $x<1$ in front of the last term within the square brackets in Eq.~\eqref{Eq:xi1}. For our computations, we set $x = 0.85$, which is a typical value for individual massive galaxies \cite{Gonzalez:2013awy} and also roughly corresponds to the cosmic average \cite{Planck:2018vyg}. 

The results are presented in Table~\ref{Tab:Constraints2}, indeed showing a small degradation of the constraints on $\Theta_0$ and $\Gamma_0$ of roughly 18\%. On the other hand, the bounds on $\mu_0$ are not significantly affected.


\section{Appendix B: Magnification and evolution biases} \label{app:mag_bias}

Galaxy surveys are usually flux limited, i.e.\ they detect only galaxies with a flux above a given threshold~$F_*$. This generates additional fluctuations in the galaxy number counts, which have been calculated e.g.\ in \cite{Bonvin:2013ogt, Challinor:2011bk}. Here we derive how this effect impacts two populations of galaxies. 

We denote by $N_\B(z,\bn)$ the number of bright galaxies per pixel, i.e.\ the number of galaxies above a chosen flux limit $F_{\rm cut}$:
\begin{align}
 N_\B(z,\bn)\equiv N(z,\bn,F\geq F_{\rm cut}) \, . 
\end{align}
Since light propagation is affected by inhomogeneities, $F_{\rm cut}$ corresponds to a different luminosity threshold in different directions: $L_{\rm cut}(z,\bn)=\bar{L}_{\rm cut}(z)+\delta L_{\rm cut}(z,\bn)$. Here, $\bar{L}_{\rm cut}$ denotes the luminosity threshold associated to $F_{\rm cut}$ in a homogeneous Universe, and $\delta L_{\rm cut}$ is the departure from this average due to fluctuations.
We obtain
\begin{align}
&N_\B(z,\bn)=N\left(z,\bn,L\geq \bar{L}_{\rm cut}(z)+\delta L_{\rm cut}(z,\bn)\right) \nonumber\\
&\simeq N\left(z,\bn,L\geq \bar{L}_{\rm cut}(z)\right)-\frac{5}{2}s(z,L_{\rm cut})\frac{\delta L_{\rm cut}}{\bar{L}_{\rm cut}}\, , \label{eq:NB}
\end{align}
where 
\begin{align}
s(z,L_{\rm cut})\equiv -\frac{2}{5}\frac{\partial}{\partial L_{\rm cut}}N(z,L\geq L_{\rm cut})\, .
\end{align}
On the other hand, the faint galaxies have a flux smaller than $F_{\rm cut}$, but larger than the flux threshold of the survey~$F_*$:
\begin{align}
 N_\F(z,\bn)&\equiv N(z,\bn,F_{\rm cut}> F\geq F_*)\nonumber\\
 &=N(z,\bn,F\geq F_*)-N(z,\bn,F\geq F_{\rm cut})\nonumber\\
 &\simeq N\left(z,\bn,\bar{L}_{\rm cut} > L\geq \bar{L}_*\right)\nonumber\\
 &\phantom{\simeq}-\frac{5}{2}s(z,L_{*})\frac{\delta L_*}{\bar{L}_*}+\frac{5}{2}s(z,L_{\rm cut})\frac{\delta L_{\rm cut}}{\bar{L}_{\rm cut}}\, ,\label{eq:NF}
\end{align}
where 
\begin{align}
s(z,L_*)\equiv -\frac{2}{5}\frac{\partial}{\partial L_*}N(z,L\geq L_*)\, .
\end{align}

\begin{figure}
    \centering
    \includegraphics[width=0.48\textwidth]{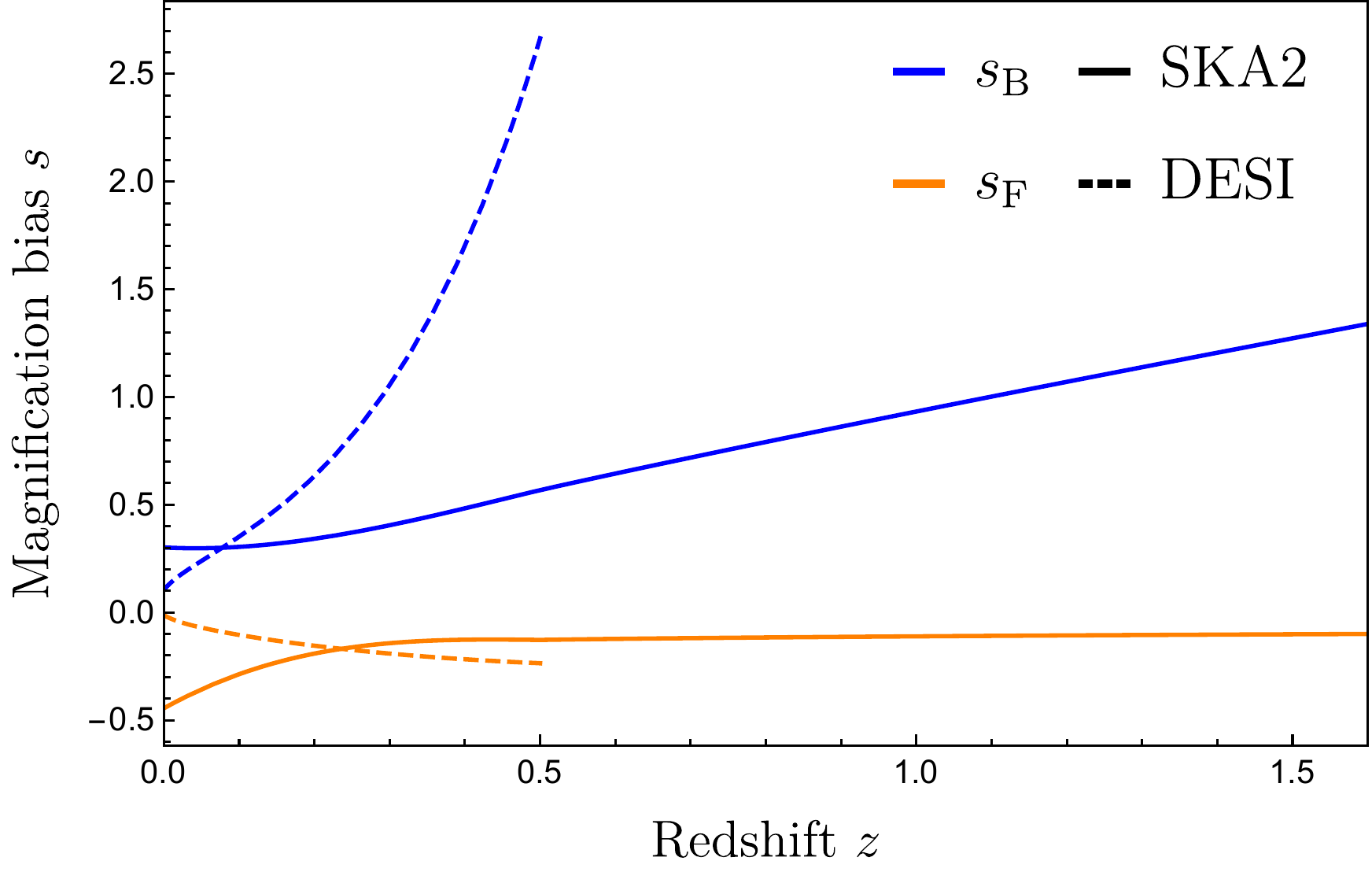}
    \caption{Magnification biases of the bright and faint populations of SKA2 and the Bright Galaxy Sample of DESI.}
    \label{fig:Magbias}
\end{figure}

For a fixed flux, the fluctuations in luminosity are directly related to the fluctuations of the luminosity distance, which have been calculated in~\cite{Bonvin:2005ps,Hui:2005nm}:
\begin{align}
\frac{\delta L_{\rm cut}}{\bar{L}_{\rm cut}}= \frac{\delta L_*}{\bar{L}_*}=2\frac{\delta d_L(z,\bn)}{\bar d_L(z)}\, .\label{eq:dL} 
\end{align}
Here we are only interested in the terms contributing to the dipole, i.e.~those proportional to the peculiar velocity. Inserting Eq.~\eqref{eq:dL} into~\eqref{eq:NB} and~\eqref{eq:NF}, we find that the flux thresholds generate fluctuations in $\Delta$ for the bright and faint populations of the form
\begin{align}
\Delta^{\rm mag}_{\B,\F}&=-5s_{\B,\F}(z) \left(1-\frac{1}{r\HH}\right)\bV\cdot\bn\,, 
\end{align}
where
\begin{align}
s_\B(z)&\equiv s(z,L_{\rm cut})\,,  \\
s_\F(z)&\equiv s(z,L_{*})-s(z,L_{\rm cut})\,.
\end{align}
To calculate $s_\B(z)$ and $s_\F(z)$ for SKA2, we use the fitting function for $s(z,L)$ given in~\cite{Camera:2014bwa}, with 
a flux sensitivity limit $F_*$ of $5\,\mu$Jy. We then choose the flux cut $F_{\rm cut}$ in each redshift bin such that we have the same number of bright and faint galaxies. For DESI, we use the model developed in~\cite{Jelic-Cizmek:2020pkh} for the magnification bias of the BGS with a magnitude limit $m_*=19.5$, and we set the magnitude cut by again imposing the same number of bright and faint galaxies. The resulting magnification bias functions for the bright and faint populations of both surveys are shown in Fig.~\ref{fig:Magbias}.~\footnote{Note that the $s_\F$ plotted in Fig.~\ref{fig:Magbias}  differs from the one in~\cite{Bonvin:2020cxp}, which is defined as $s_\F=s(z,L_*)$ and consequently wrongly neglects the impact of the flux cut $F_{\rm cut}$ on the faint population.}

The number counts fluctuations $\Delta^{\rm rel}$ also depend on the evolution bias $f^{\rm evol}$, see Eq.~\eqref{eq:Delta_rel}. The evolution bias describes the evolution of the galaxy population with time, while taking the selection function into account~\cite{Challinor:2011bk}. For the forecasts, we use $f^{\rm evol}_\B=f^{\rm evol}_\F=0$. Once data will be available, the magnification bias and the evolution bias can be measured from the average number of galaxies.

\bibliographystyle{apsrev4-1}
\bibliography{growth_ebreak.bib}

\end{document}